\documentclass[aps]{revtex4-2}
\usepackage{graphicx}
\usepackage{microtype}
\usepackage{amsmath}
\usepackage{amssymb}
\usepackage{subfigure}
\usepackage{hyperref}
\usepackage{url}
\usepackage{xcolor}
\usepackage{color}
\usepackage{mathrsfs}
\usepackage{amsfonts}
\usepackage{latexsym}
\usepackage{ragged2e}
\usepackage{textcomp}
\usepackage{phaistos}
\usepackage[utf8]{inputenc}
\usepackage[normalem]{ulem}
\usepackage{float}
\usepackage{natbib}

\makeatletter
\renewcommand\@makefnmark{\hbox{\@textsuperscript{\normalfont\color{purple}\@thefnmark}}}
\renewcommand\@makefntext[1]{%
  \parindent 1em\noindent
            \hb@xt@1.8em{%
                \hss\@textsuperscript{\normalfont\@thefnmark}}#1}
\makeatother

\usepackage{caption}
\DeclareCaptionJustification{justified}{\leftskip=0pt \rightskip=0pt \parfillskip=0pt plus 1fil}
\definecolor{vividviolet}{rgb}{0.62, 0.0, 1.0}
\definecolor{amaranth}{rgb}{0.9, 0.17, 0.31}
\definecolor{palatinateblue}{rgb}{0.15, 0.23, 0.89}
\definecolor{brightpink}{rgb}{1.0, 0.0, 0.5}
\definecolor{cornflowerblue}{rgb}{0.39, 0.58, 0.93}
\definecolor{deepcarminepink}{rgb}{0.94, 0.19, 0.22}
\definecolor{radicalred}{rgb}{1.0, 0.21, 0.37}

\hypersetup{ linktoc=all,
    colorlinks, linkcolor={palatinateblue},
    citecolor={brightpink}, urlcolor={amaranth}
}

\begin{document}

% ============================================================
%  TITLE AND AUTHORS
% ============================================================
\title{Geometric Capture and Optical Critical Curves of Purely Electric, Magnetic and Dyonic Black Holes in Einstein–Euler–Heisenberg Theory }
\author{Yong-Zhuang Li}
\email{liyongzhuang@just.edu.cn}
\affiliation{School of Science, Jiangsu University of Science and Technology, 212100, Zhenjiang, China.}
\author{An-Hao Xu}
\email{yangmomota336@gmail.com}
\affiliation{School of Science, Jiangsu University of Science and Technology, 212100, Zhenjiang, China.}
\author{Haoyuan Luo}
\email{2770521341@qq.com}
\affiliation{School of Science, Jiangsu University of Science and Technology, 212100, Zhenjiang, China.}
\author{Xiao Yan Chew }
\email{xiao.yan.chew@just.edu.cn}
\affiliation{School of Science, Jiangsu University of Science and Technology, 212100, Zhenjiang, China.}

\begin{abstract}
We examine background null capture in the electric, magnetic, and dyonic solutions of Einstein gravity coupled to a quartic Euler–Heisenberg-type electromagnetic Lagrangian. The calculation uses the exact hypergeometric electric solution, the closed-form magnetic solution, and a numerical dyonic solution. Relative to the Reissner--Nordstr\"om value at fixed total charge, the EEH correction increases the geometric critical impact parameter in the parameter ranges considered. For the purely magnetic case, the disappearance of the two outer horizons can expose an additional capture barrier and thus forms an unstable circular orbit with a very small radius.  In the dyonic case, for a fixed magnetic-to-electric charge ratio, the behavior is consistent with that of the pure electric solution while preserving the phase transition between single- and multi-horizon black hole phases. Moreover, when the coupling parameter is fixed, the effect of the magnetic-to-electric charge ratio depends on the electric charge. The present findings establish that the multi-horizon phase can impart pronounced signatures to geometric photon observables, whereas, in the off-transition regime, the purely electric, purely magnetic, and dyonic solutions remain characterized by clearly distinguishable trends. 

\end{abstract}

\date{\today}

\maketitle

\section{Introduction}

 Recently, black-hole imaging has become  one of the important astronomical observations. The Event Horizon Telescope (EHT) Collaboration has obtained horizon-scale images of the supermassive black holes M87* and Sagittarius~A* \cite{EventHorizonTelescope:2019dse,EventHorizonTelescope:2022wkp}, establishing black-hole shadows as powerful probes of strong-field gravity. The formation of shadow boundary by the photon sphere around the black holes could encode their geometries of spacetimes that is largely independent of the astrophysical environment, making it an ideal tool for testing deviations of different gravitational theories from general relativity (GR) and from Einstein--Maxwell (EM) theory \cite{Uniyal:2022vdu}.

It is well known that the Reissner--Nordstr\"om (RN) black hole suffers from the mass-inflation instability at its Cauchy horizon \cite{Simpson:1973ua,Poisson:1990eh,Ori:1991zz}. This instability challenges the viability of the RN geometry as a stable description of charged black holes in the strong-field regime and has motivated the search for black-hole solutions beyond linear electrodynamics. A natural framework for such generalizations is nonlinear electrodynamics (NED), which arises in the strong-field regime and produces charged black holes whose geometry may differ substantially from that of the RN solution \cite{Bronnikov:2017sgg,Sorokin:2021tge,Bronnikov:2022ofk,Lan:2023cvz,Luo:2024bdz}. Therefore, NED provides a physically  well motivated reason to investigate how  modifications on Maxwell electrodynamics  could change the geometry, internal structure, and stability of charged black holes. Among NED models, the Euler--Heisenberg (EH) effective theory \cite{Heisenberg:1936nmg,Schwinger:1951nm,Dunne:2004nc} stands out as it is derived directly from  the truncation of one-loop quantum electrodynamics (QED), providing the leading vacuum-polarization corrections to Maxwell theory. The experimental foundation for EH-type nonlinearities has been reinforced by recent polarization measurements of neutron stars and magnetars \cite{Mignani:2016fwz,Stewart:2025cmt,Taverna:2026syy},  giving observational motivation to the study of Einstein--Euler--Heisenberg (EEH) black holes. 

EEH black holes have attracted considerable theoretical attention in recent years \cite{Ruffini:2013hia,Yajima:2000kw,Amaro:2020xro,Magos:2020ykt,Guerrero:2020uhn,Kruglov:2017ymn,Breton:2021mju,Kala:2025kkm,Bakopoulos:2024hah,Liang:2025hzr}, with studies addressing their thermodynamics \cite{Gogoi:2023wih,Zhao:2024phz,Awal:2026vfu}, quasinormal modes \cite{Kala:2025kkm,Feng:2026yhj}, scalarization \cite{Zhang:2025msi,Guo:2025ksj,Guo:2026qib,Zhang:2026bqu,Guo:2026wia}, and their extension to AdS/dS backgrounds \cite{Li:2021ygi,Magos:2020ykt,Maceda:2018zim,Tan:2026itp}. Several shadow and gravitational lensing studies have also been carried out for specific EEH configurations \cite{MONDAL2025116859,Hamil:2024njs,Jafarzade:2025byr,Huang:2025jfa,Myung:2025dey,Uniyal:2022vdu,Ditta:2025ezx,Kazempour:2026fos,Huang:2026tul}. However, most of these works focus on a single family of EEH solutions or employ simplified approximate metric functions. A systematic shadow analysis encompassing the complete family of static, spherically symmetric EEH black holes, including purely electric, purely magnetic, and dyonic solutions, remains absent. Besides, whether rich horizon structures can arise naturally within a physically motivated effective theory such as EH electrodynamics also remains an open question. Recently, Luo \textit{et al.} have constructed three families of EEH black holes within a unified framework \cite{Luo:2026eeh,Luo:2026ndd}. A notable result is that the purely magnetic and dyonic solutions admit a novel three-horizon phase, consisting of one event horizon and two inner horizons, for sufficiently weak EH coupling and small magnetic-to-electric charge ratio. The EH nonlinear interaction modifies the horizon structure and thermodynamic properties of charged black holes in a qualitatively new way, while the central curvature singularity persists. The three-horizon phase is a distinctive prediction of EEH theory that has no analogue in EM theory and raises the question of whether this multi-horizon structure could leave observable imprints on the black-hole shadow.

Among the most directly related optical studies, Magos~\textit{et al.}~\cite{Magos:2023nnb} considered magnetically and dyonically charged EEH black holes using a Hamiltonian description and the Pleba\'{n}ski effective pseudometric. Polarization-dependent EH propagation and magnetic critical curves were also studied in Refs. \cite{Breton:2021mju,dePaula:2026blu}.  These results establish the importance of distinguishing null geodesics of the spacetime metric from the characteristics of the electromagnetic equations. Electromagnetic characteristics contain an additional constitutive response that depends on polarization. Separating these contributions makes it possible to identify the changes caused by gravitational backreaction before incorporating the propagation properties of a particular field. 

The distinction between geometric and electromagnetic capture is particularly relevant when several circular null orbits coexist. An outer unstable orbit can govern a family of scattering trajectories without determining the capture threshold, which depends on the largest accessible potential barrier. Moreover, a formal geometric capture orbit need not lie in a domain admitting regular electromagnetic propagation. These considerations motivate a joint analysis of orbit selection, characteristic propagation, and the local validity of the nonlinear-electrodynamic description. In this work, we examine these issues for the purely electric, magnetic, and dyonic EEH solutions. 

The remainder of this paper is organized as follows. In Sec.~\ref{sec:frame}, we introduce the EEH theory and present the three metric functions $N(r)$ that constitute the basis of our analysis. In Sec.~\ref{sec:photon}, we presents the null-orbit and strong-deflection analysis, compares the three charge sectors, and discusses geometric degeneracies and orbital instability.  We conclude in Sec.~\ref{sec:Conc} with a summary of our results and an outlook on future directions.

\section{Theoretical framework}
\label{sec:frame}

The action of the EEH theory is described by 
\begin{equation}
S=\frac{1}{16\pi}\int d^4x\,\sqrt{-g}\left[R-\mathcal{L}_{\rm EH}(\mathcal{F},\mathcal{G})\right]\,,
\end{equation}
with the one-loop EH effective Lagrangian~\cite{Heisenberg:1936nmg}
\begin{equation}
\mathcal{L}_{\rm EH}(\mathcal{F},\mathcal{G})
=\mathcal{F}-\epsilon\left(\mathcal{F}^2+\frac{7}{4}\mathcal{G}^2\right)\,,
\end{equation}
where $\epsilon$ is the EH nonlinear coupling parameter. The two electromagnetic invariants are
\begin{equation}
\mathcal{F}\equiv F_{\mu\nu}F^{\mu\nu}\,,\qquad
\mathcal{G}\equiv \tilde{F}^{\mu\nu}F_{\mu\nu}\,,
\end{equation}
where $F_{\mu\nu}=\partial_\mu A_\nu-\partial_\nu A_\mu$ is the electromagnetic field tensor, $\tilde{F}^{\mu\nu}=\frac{1}{2}\eta^{\mu\nu\rho\sigma}F_{\rho\sigma}$ is its Hodge dual, and $\eta^{\mu\nu\rho\sigma}$ is the completely antisymmetric Levi-Civita tensor. In the limit $\epsilon=0$, the theory reduces to the standard EM theory.

Three families of EEH black holes have been constructed in the companion work~\cite{Luo:2026eeh}, which  are obtained from the most general static, spherically symmetric gauge potential compatible with electric and magnetic charges $(Q, P)$. The spacetime metric takes the form
\begin{equation}
\label{eq:metric}
ds^2=-e^{-2\sigma(r)}N(r)dt^2+\frac{dr^2}{N(r)}+r^2\left(d\theta^2+\sin^2\theta\,d\varphi^2\right)\,,
\end{equation}
with $N(r)=1-2m(r)/r$, where $m(r)$ is the Misner--Sharp mass function and the ADM mass is $M=\lim_{r\rightarrow\infty}m(r)$. The Einstein equations imply $N(r)\sigma'(r)=0$, so that $\sigma(r)$ is constant outside any horizons.  Asymptotic flatness together with a rescaling of the time coordinate allow one to set $\sigma(r)=0$~\cite{Luo:2026eeh}. Consequently, the spacetime is fully determined by the single metric function $N(r)$.

For purely electric ($P=0$) or purely magnetic ($Q=0$) configurations one has $\mathbf{E}\cdot\mathbf{B}=0$ and therefore $\mathcal{G}=0$, suggesting the $\mathcal{G}^2$ term in the Lagrangian becomes indispensable only for dyonic configurations. The cubic equation governing $E(r)$ has discriminant $\Delta>0$ for $\epsilon>0$, so that a unique real electric-field branch exists for all physically relevant parameters. Meanwhile, it is convenient to work with dimensionless quantities. Under the scaling symmetry of the EEH field equations, $r\rightarrow\lambda r$, $(M,Q,P)\rightarrow\lambda(M,Q,P)$, $\epsilon\rightarrow\lambda^2\epsilon$, the combinations
\begin{equation}
{\epsilon}\rightarrow\frac{\epsilon}{M^2}\,,\qquad
Q\rightarrow\frac{Q}{M}\,,\qquad
P\rightarrow\frac{P}{M}\,,\qquad
\beta\rightarrow\frac{P}{Q}
\end{equation}
remain invariant. Here $Q$ and $P$ are the electric and magnetic charge-to-mass ratios now, while $\beta$ parametrizes the relative strength of the magnetic and electric charges. Throughout the shadow analysis we set $M=1$.

Since the shadow geometry is determined entirely by the metric function $N(r)$, we summarize the three solutions to be analyzed. Their derivations and full properties are presented in the companion work~\cite{Luo:2026eeh}. Here we collect only the expressions for $N(r)$ relevant to the photon-sphere and shadow computations.

\noindent\textbf{$\bullet$\;Purely electric case \;$(P=0)$.}\quad
The companion work~\cite{Luo:2026eeh} provides an exact analytical solution in terms of Gaussian hypergeometric functions ${}_2F_1$; for the shadow analysis, we evaluate this exact solution directly at the photon sphere. For reference we also display the asymptotic expansion, which reduces to the RN metric at leading order:
\begin{equation}
\label{eq:N_electric}
N(r)=1-\frac{2m(r)}{r},
\end{equation}
where 
\begin{equation}
    m(r)=M+U_{1}(u)\left[U_{2}(u)-8\left({}_{2}F_{1}\left(\frac{1}{2},\frac{7}{12};\frac{3}{2};1-u\right)+ {}_{2}F_{1} \left(\frac{1}{2},\frac{11}{12};\frac{3}{2};1-u\right)\right)\right].
\end{equation}
The dimensionless variable $u$ is defined by $z^2=r^2/\sqrt{27\epsilon Q^2}$ as $u=\left(\sqrt{z^4+1}-1\right)/\left(\sqrt{z^4+1}+1\right)$. $U_{1}$ and $U_{2}$ are defined as 
\begin{align}
    U_{1}(u)&=\frac{Q^{3/2}\sqrt{2(1-u)}}{24\sqrt[4]{27} \epsilon^{1/4}};\\
    U_{2}(u)&=\frac{3\left(11+8u^{1/3}+11u^{2/3}\right)u^{1/12}}{(1+u^{1/3}+u^{2/3})^2}.
\end{align}

\noindent\textbf{$\bullet$\;Purely magnetic case \;$(Q=0)$.}\quad
There are two solutions for this case with different asymptotic boundaries. The asymptotically flat magnetic solution  is given in closed form:
\begin{equation}
\label{eq:N_magnetic}
N(r)=1-\frac{2M}{r}+\frac{P^2}{r^2}-\frac{2\epsilon P^4}{5r^6}\,.
\end{equation}
The term proportional to $\epsilon$ originates from the EH correction. For sufficiently weak coupling $\epsilon\lesssim0.076$  and restricted $P$, the horizon equation $N(r)=0$  admits up to three positive real roots, i.e., one event horizon and two inner horizons, which is a novel multi-horizon phase absent in EM theory. 

\noindent\textbf{$\bullet$\;Dyonic case \;$(Q\neq0,P\neq0)$.}\quad
The dyonic solutions are constructed numerically in the companion work which has an asymptotic expansion
\begin{equation}
\label{eq:N_dyonic}
N(r)=1-\frac{2M}{r}+\frac{(1+\beta^2)Q^2}{r^2}
-\frac{2\epsilon(1+5\beta+\beta^2)Q^4}{5r^6}
+\frac{4\epsilon^2(2+5\beta^2)^2Q^6}{9r^{10}}+O(r^{-14})\,.
\end{equation}
For $\epsilon=0$ the first three terms reproduce the dyonic RN solution, while the limit $\beta=0$ recovers Eq.~\eqref{eq:N_electric}. The dyonic family continuously interpolates between the purely electric and purely magnetic limits. One can cross-check that the expansion agrees with the full numerical solution of Ref.~\cite{Luo:2026eeh} to within $10^{-4}$ in $N(r_{\rm ph})$ for validated parameter region ($Q, \epsilon, \beta$). However, we focus on the numerical solution in the following calculations.

\section{NULL GEODESICS, CAPTURE, AND STRONG DEFLECTION }
\label{sec:photon}

\subsection{Null effective potential and critical impact parameters }
\label{sec:photon_null}

For the metric~\eqref{eq:metric}, the null geodesics possess the conserved quantities
\begin{equation}
E=N(r)\dot t\,,\qquad
L=r^2\dot\varphi\,,
\end{equation}
where a dot denotes differentiation with respect to an affine parameter, $E$ and $L$ are the energy and angular momentum of a photon, respectively.  By spherical symmetry,  one can just study the motion  of a photon  on  the equatorial plane $(\theta=\pi/2)$. The radial equation reduces to
\begin{equation}
\label{eq:radial}
\dot r^2=E^2-\frac{L^2}{r^2}N(r)\equiv E^2-V_{\rm eff}(r)\,,
\end{equation}
with the effective potential
\begin{equation}
V_{\rm eff}(r)=\frac{L^2}{r^2}N(r)\,.
\end{equation}
The effective potential governs the qualitative behavior of null geodesics and determines both the photon sphere and the shadow boundary. For a fixed angular momentum $L$, $V_{\rm eff}(r)$ exhibits a single local maximum outside the event horizon, which locates the unstable circular photon orbit $r_{\rm ph}$. Photons arriving from infinity with impact parameter $b\equiv L/E$ slightly larger than the critical value $b_c=R_{\rm sh}$ are deflected away from the black hole after reaching a turning point $r_0>r_{\rm ph}$, while photons with $b<b_c$ cross the photon sphere and are captured.

Clearly, the EH nonlinear coupling modifies the effective potential through the correction terms in $N(r)$. In all three families, the leading EH correction enters $N(r)$ with a \emph{negative} coefficient, which lowers the effective potential $V_{\rm eff}(r)=L^2N(r)/r^2$ in the photon-sphere region. This reduction shifts the local maximum of $V_{\rm eff}$ to larger radii, suggesting  that the photon sphere expands while simultaneously lowering the barrier height $V_{\rm eff}(r_{\rm ph})$ where $r_{ph}$ is the photon sphere determined by 
\begin{equation}
\label{eq:photosphere}
\frac{d}{dr}\left(\frac{N(r)}{r^2}\right)=0.
\end{equation}
For each of the three families of solutions, one can easily verify numerically that Eq.~\eqref{eq:photosphere} admits a unique real root outside the event horizon throughout the parameter regions shown in the figures. This root is the geometric capture threshold. Furthermore, for an unstable circular orbit, the associated geometric critical impact parameter is 
\begin{equation}
\label{eq:shadow}
R_{\rm sh}=b_c=\frac{r_{\rm ph}}{\sqrt{N(r_{\rm ph})}}\,.
\end{equation}
We should make it clear that the notation $R_{\text{sh}}$ in the above equation and the retained figure axes denotes the geometric critical impact parameter $b_c$, not a polarization-dependent electromagnetic shadow radius. In a single-barrier exterior this equals the capture threshold. When several unstable orbits coexist, each has its own critical impact parameter and the global capture criterion below must be applied separately.

\subsubsection{Geometric vs.\ pseudometric photon propagation}
\label{sec:geom_pseudo}

Magos et al. \cite{Magos:2023nnb} studied neutral and charged particle trajectories in magnetic and dyonic EEH spacetimes and used a Pleba\'nski pseudometric to calculate photon trajectories and shadows. Their treatment distinguishes background null geodesics from electromagnetic propagation, as does the present analysis. The two studies also share the same leading nonlinear correction to the background geometry. This agreement can be made explicit by matching the charge and coupling conventions. The magnetic metric function in Eq. (24) of Ref. \cite{Magos:2023nnb} is 
\begin{equation}
    f_m(r)=1-\frac{2M}{r}+\frac{Q_{m}^2}{r^2}-\frac{\alpha_{f} Q_m^{4}}{225\pi E_{c}^2r^6},
\end{equation}
where $\alpha_{f}$ is the fine-structure constant and $E_c$ is the critical electric field in the conventions of Ref. \cite{Magos:2023nnb}. With the same normalization of the charge and units, the identification
\begin{equation}
    P=Q_m,\quad \epsilon=\frac{\alpha}{90\pi E_c^2}
\end{equation}
reproduces to Eq. (\ref{eq:N_magnetic}). So the screened magnetic charge used in that reference can be seen as a radius-dependent rewriting of this metric. Expanding the corresponding dyonic expression, i.e., Eqs. (114)–(116) of Ref. \cite{Magos:2023nnb}, gives 
\begin{equation}
    f_d(r)=1-\frac{2M}{r}+\frac{q^2}{r^2}-\frac{2\epsilon \mathcal{D}}{5r^6}+O(\epsilon^2),
\end{equation}
where $q^2=P^2+Q^2$ and $\mathcal{D}=Q^4+5 Q^2P^2+P^4$. This is the first-order expansion of Eq. (\ref{eq:N_dyonic}),  and both backgrounds reduce to RN as $\epsilon \rightarrow 0$.  The Lagrangian (F) and Hamiltonian (P) formulations describe the same constitutive theory wherever their Legendre transformation is invertible and the truncation order is matched. They do not, by themselves, define different photon propagation laws.  

The additional electromagnetic response is polarization dependent \cite{Breton:2021mju,dePaula:2026blu}. For radial fields, the characteristic metrics can be represented through $O(\epsilon)$ by

\begin{equation}\label{eq:green_optical}
\begin{aligned}
d\bar s_\pm^2&=-\mathcal A_\pm dt^2+\mathcal B_\pm dr^2+\mathcal C_\pm d\Omega^2,\\
\mathcal A_\pm&=\frac{N}{A_\pm+k_\pm\epsilon E^2},\qquad \\
\mathcal B_\pm&=\frac{1}{N(A_\pm+k_\pm\epsilon E^2)},\qquad \\
\mathcal C_\pm&=\frac{r^2}{A_\pm-k_\pm\epsilon B^2},
\end{aligned}
\end{equation}
where $A_+=1+5\epsilon\mathcal F$, $A_-=1-2\epsilon\mathcal F$, $k_+=14$, and $k_-=8$. Expanding $(\mathcal A_\pm/\mathcal C_\pm)'=0$ and $b_\pm^2=\mathcal C_\pm/\mathcal A_\pm$ about the outer RN orbit gives, at fixed $M,Q,P$,

\begin{align}
\delta r_{\rm g}&=\frac{8\epsilon\mathcal D}{5\hat{r}_{0}^4(2\hat{r}_{0}-3M)}+O(\epsilon^2),\notag\\
\frac{b_{\rm g}}{\hat{b}_0}&=1+\frac{\epsilon\mathcal D}{5\hat{r}_{0}^6\hat{N}_0}+O(\epsilon^2),\qquad\\
\frac{b_\pm-b_{\rm g}}{\hat{b}_0}&=\frac{k_\pm\epsilon q^2}{2\hat{r}_{0}^4}+O(\epsilon^2),\label{eq:green_shift}\\
\frac{b_+-b_-}{\hat{b}_0}&=\frac{3\epsilon q^2}{\hat{r}_{0}^4}+O(\epsilon^2),
\label{eq:polarization_splitting}
\end{align}
with $\hat{r}_{0}=(3M+\sqrt{9M^2-8q^2})/2$, $\hat{N}_{0}=1-2M/\hat{r}_{0}+q^2/\hat{r}_{0}^2$, and $\hat{b}_{0}=\hat{r}_{0}/\sqrt{\hat{N}_{0}}$ denote the zeroth-order RN values at fixed mass and charges. The geometric correction ($\delta r_g$, $b_g$) depends on $\mathcal D$, while the optical shift depends on $q^2$ and need not be smaller.

These expressions require a nondegenerate RN-connected orbit, $\hat{N}_{0}>0$, and small local corrections. Their positive optical shifts differ from Eqs.~(104) and (143) of Ref.~\cite{Magos:2023nnb}, so agreement of the backgrounds does not establish quantitative agreement of electromagnetic shadows. This difference cannot be assigned to the $F/P$ choice alone. The capture and strong-deflection results below concern background null geodesics. Their electromagnetic extension requires regular characteristic metrics along the rays and, for a QED interpretation, fields within the validity regime specified in Sec.~\ref{sec:frame}.

\subsubsection{Deflection angle and strong-field lensing formalism}

For a photon with impact parameter $b$, the deflection angle in a static, spherically symmetric spacetime is given by \cite{Perlick:2021aok,Ovgun:2025mil}
\begin{equation}
\label{eq:deflection}
\alpha(b)=2\int_{r_0}^{\infty}\frac{b\,dr}{r\sqrt{r^2-N(r)b^2}}-\pi\,,
\end{equation}
where $r_0$ is the turning point determined by $r_0^2=N(r_0)b^2$.
The integral~\eqref{eq:deflection} is evaluated numerically for each family of solutions over a range of impact parameters $b>b_c$. In the weak-field limit $b\gg b_c$, the deflection angle reduces to the familiar Einstein form $\alpha\approx4M/b$, with charge and nonlinear corrections at higher orders in the inverse impact parameter. As $b$ approaches the critical impact parameter $b_c$ from above, the photon executes multiple loops around the photon sphere and $\alpha(b)$ diverges logarithmically.

For the purely electric family, the exact hypergeometric solution is used; for the dyonic family, the asymptotic expansion~\eqref{eq:N_dyonic} is sufficient for evaluating the integral, since the turning point $r_0$ lies well outside the event horizon for all $b>b_c$. For the purely magnetic family, the closed-form metric function~\eqref{eq:N_magnetic} is used directly.

In the strong-field limit $b\rightarrow b_c$, the deflection angle admits the Bozza expansion \cite{Bozza:2002zj}
\begin{equation}
\label{eq:bozza}
\alpha(b)=-\bar a\ln\left(\frac{b}{b_c}-1\right)+\bar b+O(b-b_c)\,.
\end{equation}
The coefficient $\bar a$ is determined entirely by the metric function and its second derivative at the photon sphere,
\begin{equation}
\bar a=\sqrt{\frac{2}{2N(r_{\rm ph})-r_{\rm ph}^2 N''(r_{\rm ph})}}\,,
\end{equation}
whereas $\bar b$ depends on the full metric through a regularized integral. For the Schwarzschild metric, $\bar a=1$ and $\bar b\approx-0.4002$.

\subsection{Purely electric case}
\label{sec:photon_electric}

\begin{figure}[t]
\centering
\includegraphics[width=0.85\textwidth]{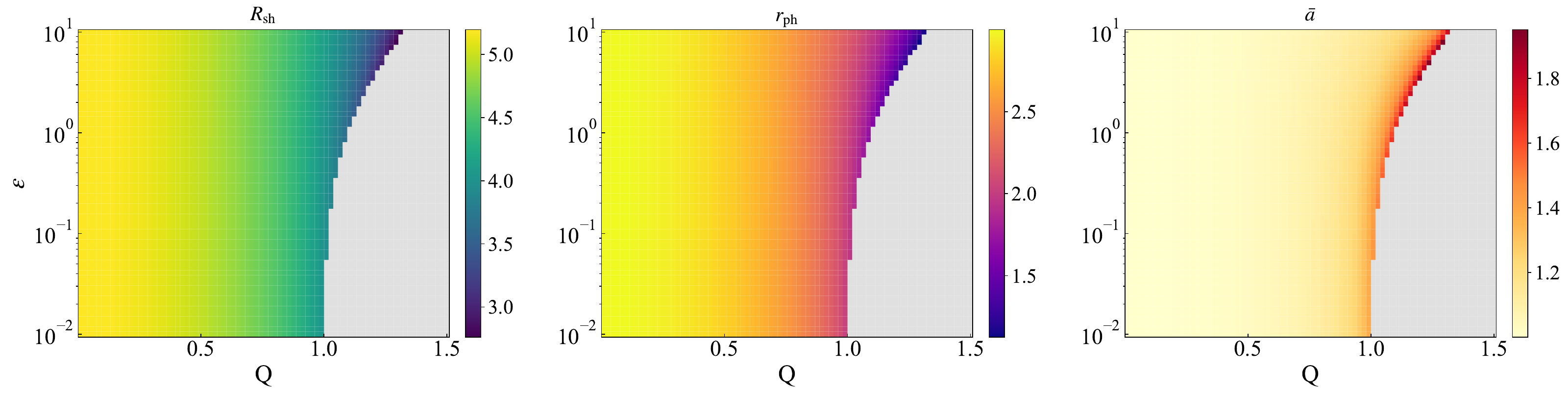}
\includegraphics[width=0.85\textwidth]{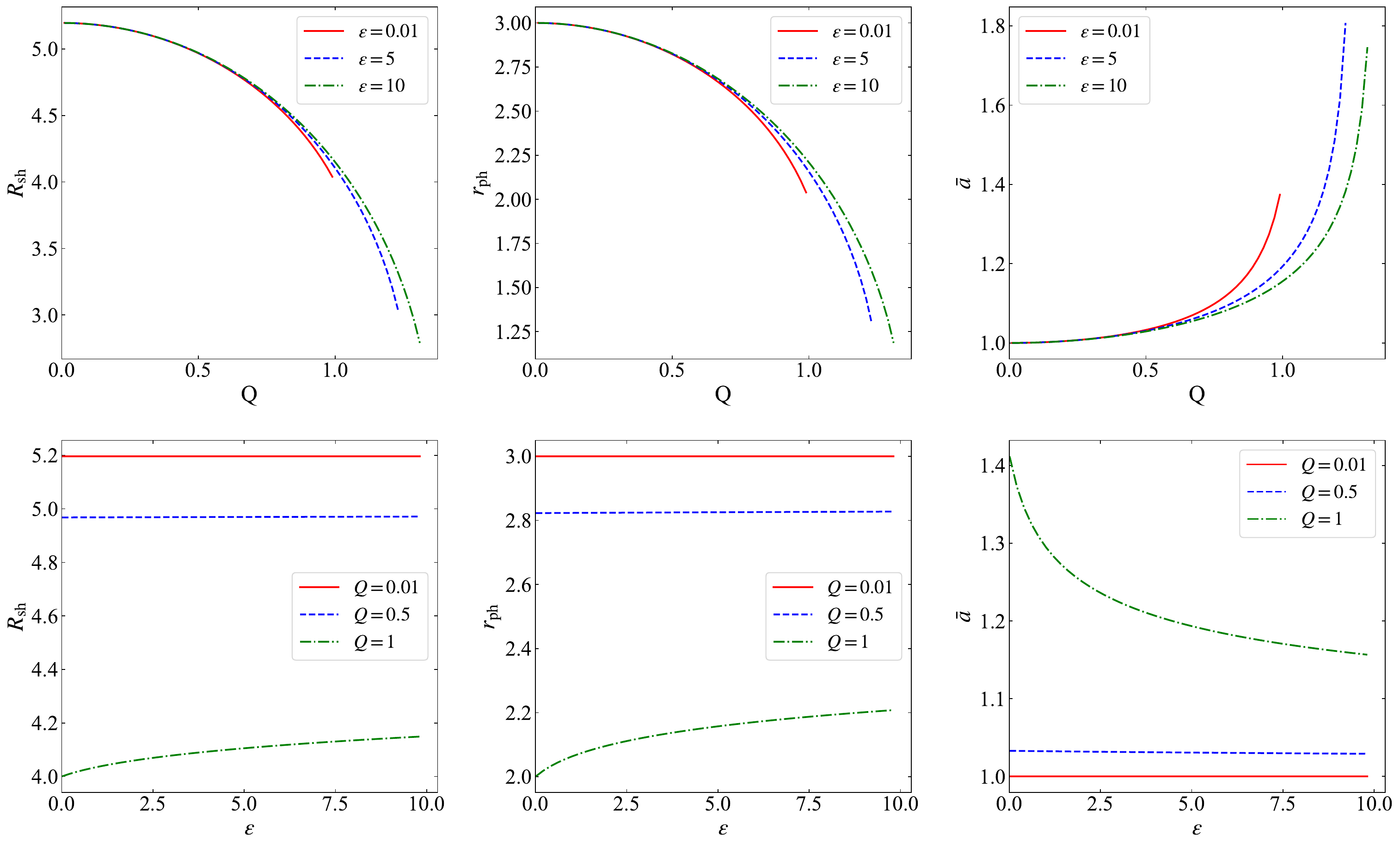}
\caption{
The influences of $\epsilon$ and $Q$ on the shadow radius $R_{\rm sh}$, photon sphere $r_{\rm ph}$ and $\bar{a}$  for purely electric EEH black holes. The gray parts in the first row show the naked singularity region.}
\label{fig:s_e_Rsh}
\end{figure}

The effects of $\epsilon$, $Q$ on the shadow radius $R_{\text{sh}}$, the photon sphere $r_{\text{ph}}$ and  the coefficient $\bar{a}$.  are displayed in Fig.~\ref{fig:s_e_Rsh} in the purely electric case.

As the charge $Q$ increases from zero, the shadow radius $R_{\rm sh}$ and $r_{\text{ph}}$ both decrease monotonically from the Schwarzschild value for all values of $\epsilon$, qualitatively similar to the RN behavior. The EH nonlinear coupling produces a small but systematic {increase} of the shadow radius relative to the RN value at equal charge. The physical origin of this increase is that the negative $-\epsilon Q^4/r^6$ correction in $N(r)$ lowers the metric function in the photon sphere region. This pushes the photon sphere outward to larger radii (where $N(r)$ is larger) and simultaneously reduces $N(r_{\rm ph})$, both of which contribute to a larger $R_{\rm sh}=r_{\rm ph}/\sqrt{N(r_{\rm ph})}$. In contrast, for the strong-field lensing case, the coefficient $\bar{a}$ increases monotonically with $Q$ for fixed $\epsilon$, while it decreases monotonically with $\epsilon$ for fixed $Q$. It is also evident that the influence of $\epsilon$ becomes more significant for larger values of $Q$.
 
\subsection{Purely magnetic case}
\label{sec:photon_magnetic}

The purely magnetic case is much more complicated as a three-horizon region exit for a specified region $(\epsilon,\,P)$ where the metric function~\eqref{eq:N_magnetic} admits one event horizon and two inner horizons, and the single-horizon region for other regions, where only the event horizon survives \cite{Luo:2026eeh}. In fact, it can be observed that when $N(r)$ transitions from the single-horizon region to the three-horizon region, or vice versa, the so-called outer horizon exhibits a jump. 

However, there is no known astrophysical mechanism that can reliably increase $P/M$ on observationally relevant timescales. Large values of $P/M$ are most plausibly primordial in origin rather than acquired through subsequent evolution \cite{Zhang:2023mpc}.

\begin{figure}[t]
\centering
\includegraphics[width=0.85\textwidth]{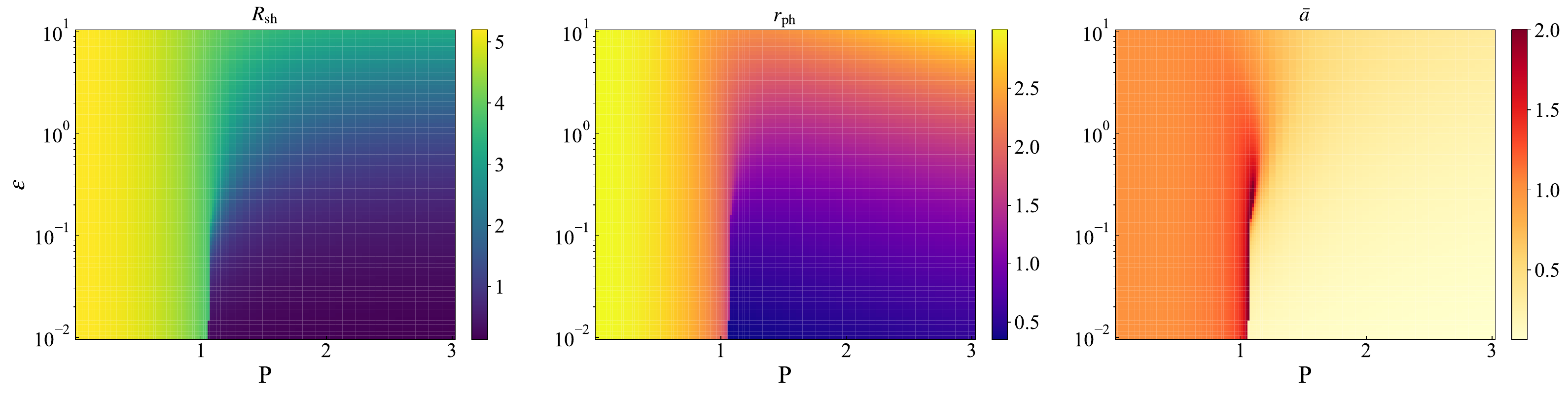}
\includegraphics[width=0.85\textwidth]{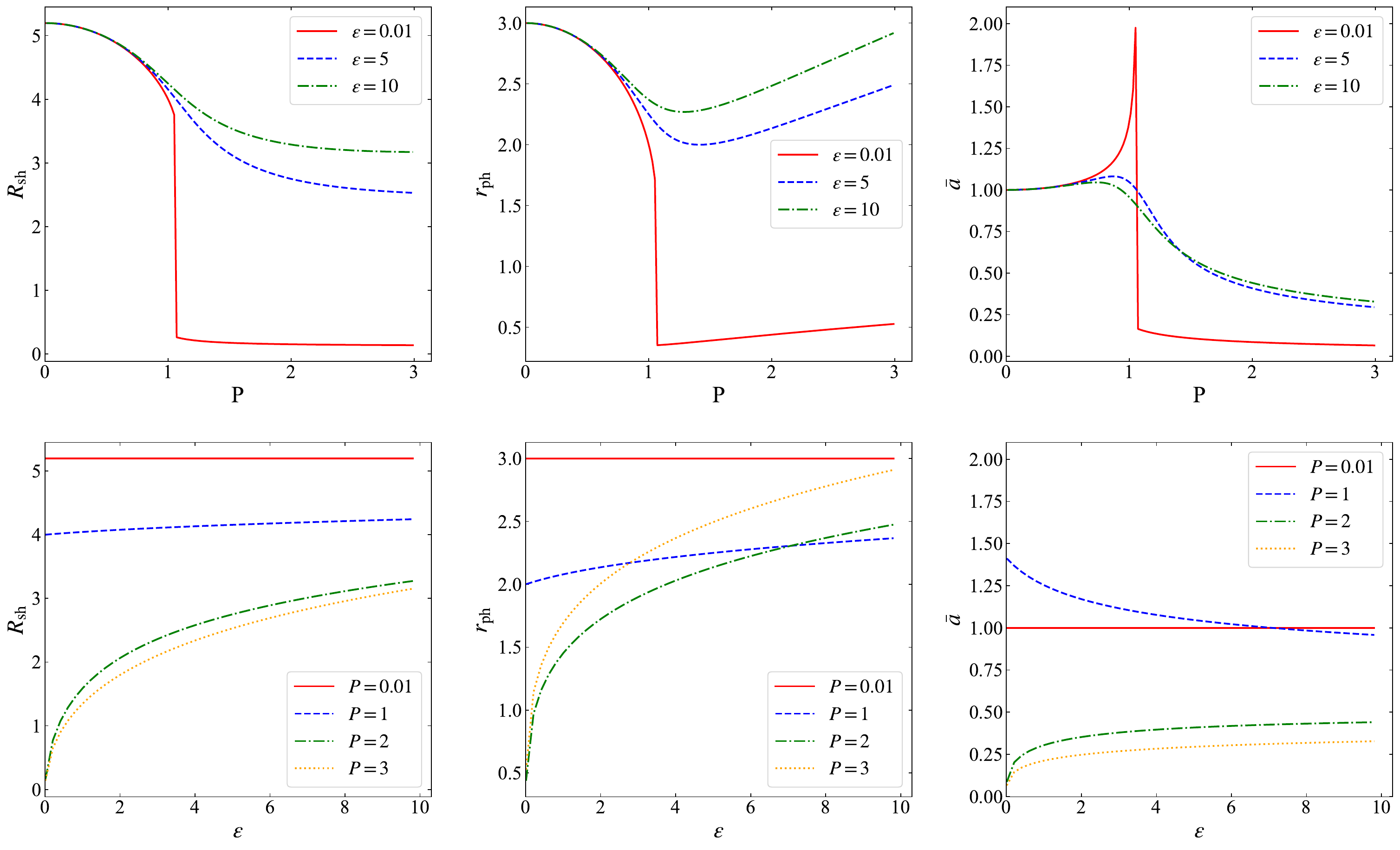}
\caption{
The influences of $\epsilon$ and $P$ on the shadow radius $R_{\rm sh}$, photon sphere $r_{\rm ph}$ and $\bar{a}$  for purely electric EEH black holes. The jump phenomenon is resulting from the small $\epsilon$ and will disappear for large enough $\epsilon$ or $\epsilon=0$.}
\label{fig:s_m_Rsh}
\end{figure}

In Fig.~\ref{fig:s_m_Rsh} we display the effects of $\epsilon$, $P$ on the shadow radius $R_{\text{sh}}$, the photon sphere $r_{\text{ph}}$ and  the coefficient $\bar{a}$. Unlike the purely electric case, the influences are not simply single-valued and monotonic. More specifically, the relations of $R_{\rm sh}$ with respect to either $\epsilon$ or $P$ are all monotonic; in particular, the influence of $\epsilon$ becomes appreciable only when $P$ is relatively large, which is a concrete manifestation of the $\epsilon$-correction term in the metric function \ref{eq:N_magnetic}. For $r_{\rm ph}$ and $\bar{a}$, however, their curves are no longer monotonic. We also note that the purely magnetic metric function, in the limit of small $\epsilon$, takes the same form as that of the RN black hole, indicating that the influence of $P$ in this case is consistent with that in the purely electric case with small $\epsilon$. Furthermore, we should particularly point out  the jump phenomenon of $R_{\rm sh}$, $r_{\rm ph}$, and $\bar{a}$ in the case of three horizons. This transition from the single-horizon phase to the three-horizon phase completely disappears when $\epsilon$ exceeds the critical value of 0.076, in which case only the single-horizon phase remains. 

Meanwhile,  within this well-defined sub-volume in the the three-horizon regime the shadow curve is nearly flat, crossing the phase boundary without any discontinuity. Physically, this establishes that the photon sphere is governed by the large-scale geometry encoded in $N(r)$ and is {decoupled} from the near-horizon causal structure that determines horizon multiplicity. This decoupling is a distinctive feature of EEH black holes with no analogue in EM theory.

While the three-horizon phase does not produce a discontinuous shadow feature, its parameter-space boundaries correlate with enhanced sensitivity of $R_{\rm sh}$ and the strong-lensing coefficient $\bar a$ to changes in the EH coupling $\epsilon$. This enhanced sensitivity provides an indirect, continuous observable proxy for the underlying multi-horizon structure, which may be exploited in parameter-estimation analyses combining shadow and strong-lensing data.

\subsection{Dyonic case}
\label{sec:photon_dyonic}

\begin{figure}[t]
\centering
\includegraphics[width=0.88\textwidth]{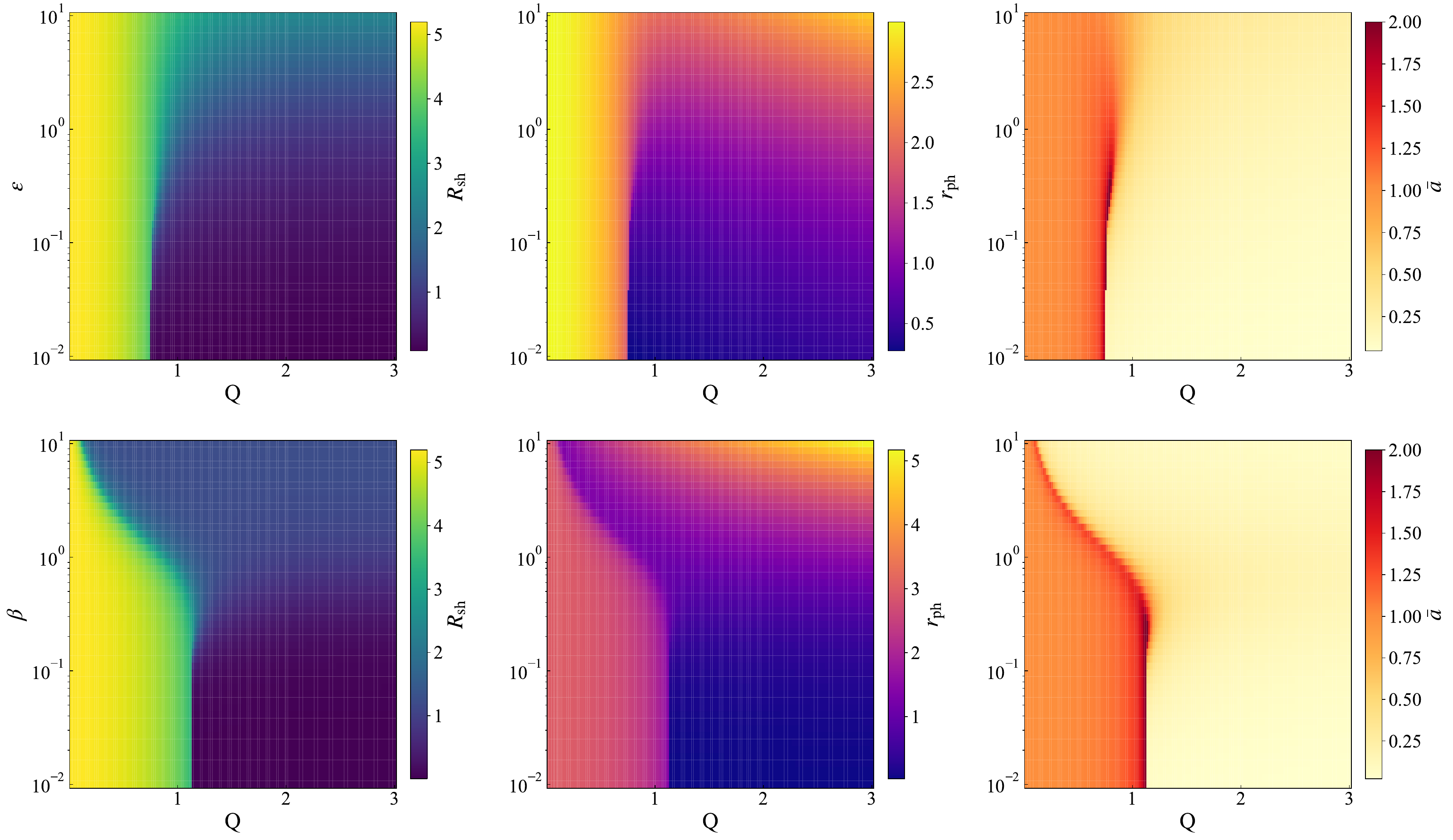}
\includegraphics[width=0.85\textwidth]{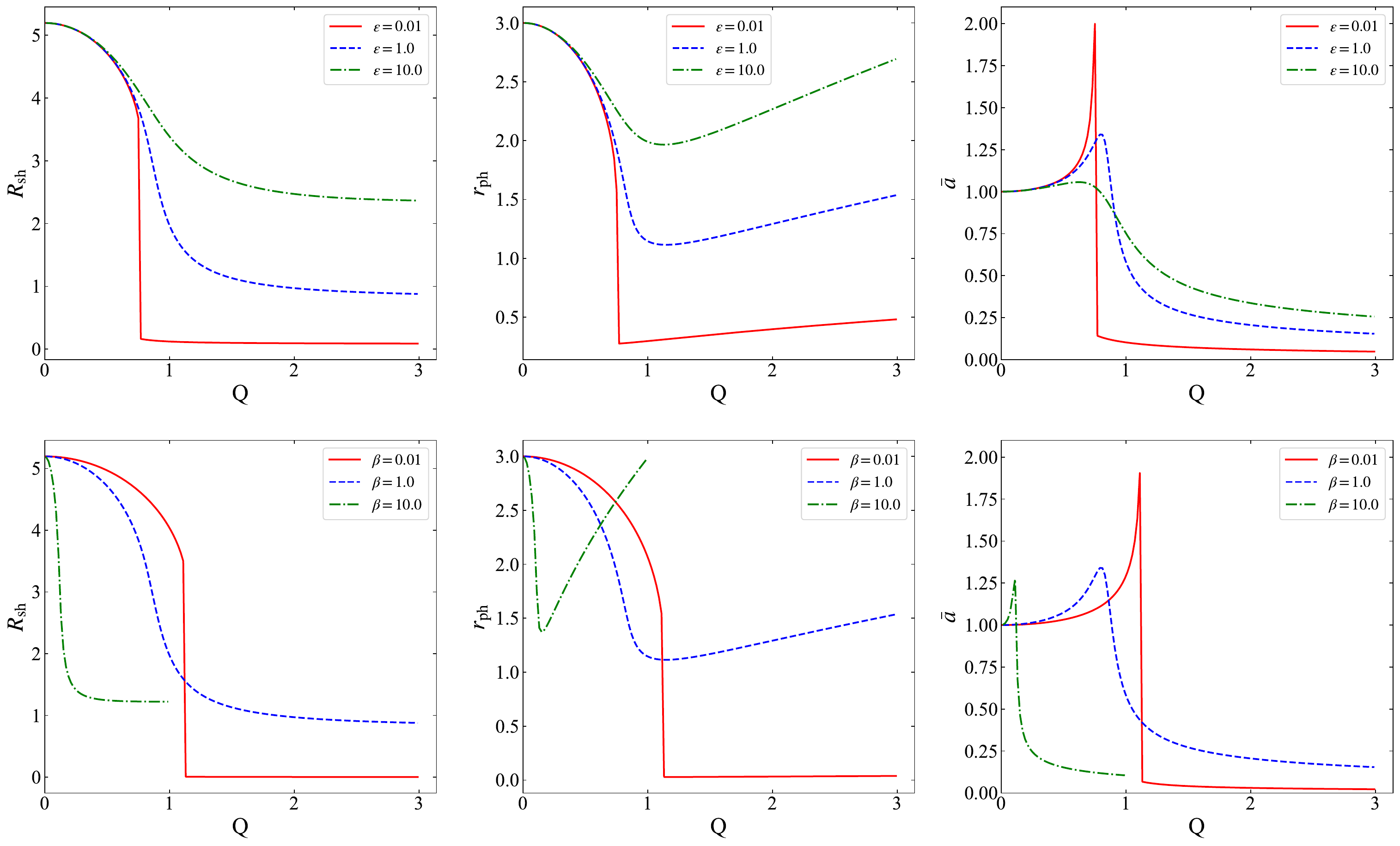}
\caption{
The influences of $\epsilon$, $\beta$ and $Q$ on the shadow radius $R_{\rm sh}$, photon sphere $r_{\rm ph}$ and $\bar{a}$  for dyonic EEH black holes.  For the density plots, we fix either $\epsilon=1$ or $\beta=1$ to investigate the effects of the parameter pairs $(\beta, Q)$ and $(\epsilon, Q)$ on the three quantities, respectively. For the line plots, with either $\beta=1$ or $\epsilon=1$ held fixed, we select distinct values for the remaining parameter ($\epsilon$ or $\beta$) to illustrate how the three quantities vary with $Q$. Note that the green dash-dotted curve in the last panel does not extend further, which is attributable to the limited numerical precision. It can be verified that for $\epsilon=1,\beta=10$, the corresponding range of $Q$ can indeed reach up to 3.  
}
\label{fig:s_d_Rsh}
\end{figure}

The dyonic case is the most general of the three families and continuously interpolates between the purely electric ($\beta\to0$) and purely magnetic ($\beta\gg1$) limits. We present density plots to illustrate the effects of the parameter pairs $(\beta, Q)$ and $(\epsilon, Q)$ on the shadow radius, photon sphere, and the coefficient, by fixing $\epsilon=1$ and $\beta=1$, respectively. Building upon these density plots, we further select discrete values of $\beta=(0.01,1,10)$ and $\epsilon=(0.01,1,10)$to investigate the dependence of the three quantities on $Q$ via curve plots. The results are displayed in Fig.~\ref{fig:s_d_Rsh}.

From the figures, we observe that when $\beta=1$ (i.e., $P=Q$) is fixed, the influences of $\epsilon$ and $Q$ on the three quantities are similar with those in the purely magnetic case. Indeed, as we initially pointed out, a smaller $\beta$ leads to behavior closer to the purely electric case, whereas a larger $\beta$ makes the purely magnetic characteristics more dominant. When $\epsilon$ is fixed, however, the situation becomes relatively more complicated. In particular, when the fixed $\epsilon$ is relatively small, the $\beta$-interval occupied by the three-horizon phase tends to become wider. 

\subsection{Discussion on degeneracy}
\label{sec:photon_degen}

The shadow radius $R_{\rm sh}$ is a single geometric observable, whereas the EEH solution is specified by the nonlinear coupling and several charge parameters.  A measured shadow size can consequently be reproduced by different combinations of $\epsilon$, $Q$, $P$, and $\beta=P/Q$.  This degeneracy is especially clear in the dyonic family as changes in the electric charge and magnetic-to-electric charge ratio can compensate one another in $R_{\rm sh}$, even though they lead to different photon-sphere radii and strong-deflection behavior, see Fig. \ref{fig:s_d_Rsh}.  The monotonic increase of $r_{\rm ph}$ and $R_{\rm sh}$ with $\epsilon$ in the purely electric case is therefore useful for interpreting that restricted family, but it does not by itself determine the EEH coupling once magnetic charge is admitted.

The strong-deflection coefficient $\bar a$ supplies complementary information because it controls the leading logarithmic behavior of the deflection angle near the critical curve.  Its nonmonotonic dependence on $\epsilon$, $P$, and $Q$ in the magnetic and dyonic sectors means that a joint measurement of the shadow scale and photon-ring structure can exclude parameter combinations that remain degenerate in $R_{\rm sh}$ alone.  The jump in $r_{\rm ph}$, $R_{\rm sh}$, and $\bar a$ at the single-to-three-horizon transition is potentially more diagnostic still.  It identifies a change in the exterior photon-sphere branch associated with a large change in the event-horizon radius.  Nevertheless, a reliable inference requires continuous tracking of the exterior branch and a comparison with the full family of permitted solutions

\subsubsection{Observational prospects}

We end the photon analysis by assessing the detectability of the EEH-induced deviations with current and forthcoming instruments as a sketchy discussion.  Horizon-scale images of M87* and Sgr~A* have already constrained the angular scale of the critical curve~\cite{EventHorizonTelescope:2019dse,EventHorizonTelescope:2022wkp}, while future very-long-baseline interferometry can improve this constraint and extract more  information about photon-ring morphology and polarization.  The most useful EEH tests would combine the shadow diameter with observables sensitive to near-critical propagation, such as the relative structure of resolved photon rings and polarization-dependent distortions.  These combination can distinguish a uniform rescaling of the critical curve from the correlated changes in $R_{\rm sh}$ and $\bar a$ found in the present parameter survey. For a photon emitted near $r_{\rm ph}$ and observed at infinity, the frequency ratio is
\begin{equation}
\frac{\nu_\infty}{\nu_{\rm ph}} = z(r_{\rm ph})\equiv\sqrt{N(r_{\rm ph})}\,,
\label{eq:redshift}
\end{equation}
which, together with the shadow radius $R_{\rm sh}=r_{\rm ph}/z$, allows the independent determination of both the photon-sphere radius $r_{\rm ph}=R_{\rm sh}\,z$ and the metric function $N(r_{\rm ph})=z^2$. This breaks the degeneracy between $r_{\rm ph}$ and $N(r_{\rm ph})$ that is inherent in the single observable $R_{\rm sh}$, providing a consistency check between the geometric and spectroscopic channels \cite{Gralla:2019phr,Frion:2021bsd}. However, astrophysical plasma surrounding the black hole introduces its own emission and absorption features that may complicate the extraction of the gravitational redshift signal; the redshift could nonetheless be isolated by fitting the continuum slope of the photon-ring spectrum, as the gravitational shift is achromatic while plasma effects are frequency-dependent. 

Finally, the photon-sphere Lyapunov exponent $\lambda_L\equiv\Omega_c/\bar a$, with the orbital angular frequency $\Omega_c=c/R_{\rm sh}$, governs the imaginary part of eikonal quasinormal-mode (QNM) frequencies via $\Im\omega_{\rm QNM}=-(n+1/2)|\lambda_L|$ \cite{Cardoso:2008bp}. Using the $R_{\rm sh}$ and $\bar a$ values numerically calculated, we find that the EEH correction modifies $\lambda_L$ by ${\sim}0.1\%$-$2\%$ relative to the RN prediction. In the $\bar a<1$ regime, $\lambda_L>\Omega_c$, i.e., stronger damping than Schwarzschild, which would manifest as a measurably faster ringdown in the gravitational-wave channel. 

\section{Summary and Discussions}
\label{sec:Conc}

We have investigated geometric null orbits and critical curves of the purely electric, purely magnetic, and dyonic black holes in the quartic Einstein–Euler–Heisenberg model.  All three families are treated within a unified computational framework: the exact hypergeometric solution for the purely electric case~\cite{Luo:2026eeh}, the exact closed-form metric for the purely magnetic case, and the numerical solution for the dyonic case.

For the purely electric solution, increasing Euler-Heisenberg parameter $\epsilon$ increases both $r_{\rm ph}$ and $R_{\rm sh}$.  The logarithmic divergence of the deflection angle near the photon sphere also becomes more rapid.  These effects are appreciable only when the electric charge is sufficiently large: for small electric charge, the $\epsilon$-induced corrections are negligible and the geometry remains close to its Reissner--Nordstr\"om counterpart.

The purely magnetic case exhibits a richer relation to the horizon structure.  For small magnetic charge $P$, varying $\epsilon$ has only a minor effect on the photon sphere and shadow.  At sufficiently large $P$, however, both $r_{\rm ph}$ and $R_{\rm sh}$ grow with $\epsilon$.  A particularly distinctive feature appears at the transition from the single-horizon to the three-horizon region: $r_{\rm ph}$, $R_{\rm sh}$, and $\bar a$ undergo a jump.  This behavior is consistent with the fact that, at the critical transition, the event-horizon radius on the three-horizon branch is much smaller than that on the single-horizon branch.  In contrast to the radii, the dependence of $\bar a$ on $\epsilon$ and $P$ in this parameter region is not monotonic.

The dyonic family combines and extends these behaviors.  With $\beta=P/Q$ fixed, the response to $\epsilon$ follows the trend of the purely electric case, namely increasing $r_{\rm ph}$ and $R_{\rm sh}$ with increasing $\epsilon$; the jump associated with the single-to-three-horizon transition nevertheless persists.  At fixed $\epsilon$, the effect of $\beta$ depends on the electric charge.  For relatively small $Q$, increasing $\beta$ decreases the photon-sphere and shadow radii, whereas for sufficiently large $Q$ the same increase enlarges both radii.  The strong-deflection coefficient is again more intricate: apart from the pronounced jump at relatively small $\beta$, $\bar a$ generally first increases and then decreases as $Q$ is increased.  Thus, neither the charge ratio nor the nonlinear coupling alone provides a monotonic control parameter for strong-field lensing in the dyonic sector.

Our results show that the multi-horizon phase can leave sharp imprints on geometric photon observables, while the purely electric, magnetic, and dyonic solutions retain clearly distinguishable trends away from the transition.  The quantities calculated here are geometric null-geodesic observables of the background metric.  Establishing their direct electromagnetic observational consequences will require a complementary analysis of the polarization-dependent optical metrics of nonlinear electrodynamics, together with ray tracing in a realistic plasma and emission model.  Such a treatment would determine which of the parameter-dependent shadow and strong-lensing signatures found here remain accessible to horizon-scale observations.

From an observational perspective, the most promising strategy is to combine several horizon-scale observables rather than infer the EEH parameters from the shadow diameter alone.  The angular shadow size constrains the critical impact parameter, whereas photon-ring substructure and the strong-deflection coefficients probe the near-critical propagation more directly.  Polarimetric information offers an additional discriminant because nonlinear electrodynamics can modify the characteristic photon cones in a polarization-dependent manner.  Joint analyses of the shadow size, ring morphology, and polarization for M87* and Sgr~A* with the EHT and its future extensions could therefore test whether the correlated variations of $R_{\rm sh}$ and $\bar a$ found here are compatible with an EEH origin.  In particular, a sharp change in these observables across the single-to-three-horizon boundary would be a distinctive target, although its practical detectability must be assessed after accounting for source variability, plasma propagation, and the uncertainties in the mass-to-distance ratio.

Several theoretical developments are needed to turn this geometric analysis into a predictive observational framework.  First, the characteristic optical metrics of the electric, magnetic, and fully dyonic F-frame solutions should be derived and used for polarization-resolved ray tracing. This will allow a direct comparison with the pseudometric treatment of nonlinear-electrodynamic photon propagation~\cite{Magos:2023nnb}.  Second, the dyonic solution should be studied beyond its present asymptotic regime through high-precision numerical constructions, with systematic root tracking across the horizon-phase boundary.  Third, coupling the resulting optical trajectories to general-relativistic magnetohydrodynamic accretion models and performing image-domain Bayesian inference would quantify which portions of the parameter space can be constrained by present and next-generation VLBI observations.  Finally, a perturbative analysis of the coupled gravitational--electromagnetic sector, including the stability of the multi-horizon branch and its quasinormal spectrum, would provide an independent gravitational-wave test of the same EEH parameters.  Together, these extensions can establish whether the geometric signatures identified here survive as robust electromagnetic and multimessenger observables.

\section*{Acknowledgment}

AX and HL are supported by the Research and Practice Innovation Plan for Graduate Students in Jiangsu Province (No.~26CXJH7332, No.~26CXJH7333). XYC is supported by National Science Foundation of China (no: W2533026).

\bibliography{mybiblio}

\end{document}